\documentstyle[sprocl]{article}

\input{psfig}

\def\be{\begin{equation}}
\def\ee{\end{equation}}
\def\bea{\begin{eqnarray}}
\def\eea{\end{eqnarray}}

\begin{document}

\phantom .
\vspace{-1cm}
{\flushright nucl-th/9712033\\}
{\flushright \ \\}

\title{NUCLEAR PHYSICS INPUT FOR SOLAR MODELS}

\author{ATTILA CS\'OT\'O}

\address{Theoretical Division, Los Alamos National Laboratory,\\
Los Alamos, NM 87545, USA; http://qmc.lanl.gov/$\sim$csoto} 


\maketitle\abstracts{We discuss microscopic cluster model descriptions of
two solar nuclear reactions, $^7{\rm Be}(p,\gamma){^8{\rm B}}$ and 
$^3{\rm He}({^3{\rm He}},2p){^4{\rm He}}$. The low-energy reaction cross
section of $^7{\rm Be}(p,\gamma){^8{\rm B}}$, which determines the
high-energy solar neutrino flux, is constrained by $^7$Be and $^8$B
observables. Our results show that a small value of the zero-energy cross
section is rather unlikely. In $^3{\rm He}({^3{\rm He}},2p){^4{\rm He}}$
we study the effects of a possible virtual state on the cross section.
Although, we have found no indication for such a state so far, its
existence cannot be ruled out yet. We calculate the $^3{\rm 
He}({^3{\rm He}},2p){^4{\rm He}}$ and $^3{\rm H}({^3{\rm 
H}},2n){^4{\rm He}}$ cross sections in a continuum-discretized coupled
channel approximation, and find a good general agreement with the data.
}

\section{Introduction}
One of the most exciting fields of research these days is neutrino
physics. Various experiments have been producing a large number of
interesting results about the neutrino, yet after decades of work we
still do not know even the most basic properties of these particles. The
pioneering experiments that measure neutrinos coming from the sun
produced the first, and still strongest, evidence for the possibility of
nonzero neutrino mass, and hence for physics beyond the standard model.
For a review of solar neutrino research, see \cite{Bahcall}. 

Solar models contain input parameters from many fields of physics. For
any reliable prediction of the solar neutrino fluxes, these parameters
must be firmly established. Nuclear physics provides the rates of solar
fusion reactions as input parameters for solar models. For a very recent
review of our current understanding of these reactions, see
\cite{Adelberger}. In the following we shall discuss a microscopic model
description of two important solar reactions, $^7{\rm
Be}(p,\gamma){^8{\rm B}}$ and $^3{\rm He}({^3{\rm He}},2p){^4{\rm
He}}$, in detail.

\section{Model and computational details}
Currently the best {\em dynamical} description of $A=6-8$ nuclei can be
achieved by the microscopic cluster model. This model assumes that the
nuclei consist of $2-3$ clusters. While the clusters are described by
simple harmonic oscillator shell-model wave functions, the intercluster
relative motions, which are the most important degrees of freedom, are
treated rigorously. This model satisfies the correct bound- and
scattering asymptotics of the wave functions, and satisfactorily
reproduces the positions of the important breakup- and rearrangement
channel thresholds and separation energies. 

Because of the low temperature of our sun (on nuclear scales) the most
effective energies for solar reactions are very low, being in the keV
region. Thus, in order to calculate charged-particle reaction cross 
sections reliably, one must use computational
methods which can supply correct bound- and scattering wave functions up
to a few hundred fermi radii with high precision. Here we use the
Kohn-Hulth\'en variational method for scattering states \cite{Kamimura} 
and the Siegert variational method for bound states \cite{Fiebig}. We
briefly discuss these methods in a simplified manner, the generalization
for realistic calculations is straightforward.

The Kohn-Hulth\'en method starts with the following trial wave function
\begin{equation}
\Psi^t=\sum_{i=1}^Nc_i\varphi_i+\phi^-_E-S(E)\phi^+_E.
\end{equation}
Here $\varphi_i$ are square-integrable functions (Gaussians in our case)
while $\phi^-_E$ and $\phi^+_E$ are incoming and outgoing Coulomb 
functions with energy $E$, respectively. From the $\left\langle\right .
\delta \Psi^t\left \vert \right .\widehat H-E \left \vert \right .
\Psi^t\left .\right\rangle=0$ projection equation one gets 
a set of linear equations that can be solved for the $c_i$
coefficients and the $S$ scattering matrix. For many-body systems one can
use basis functions that are made by matching $\varphi_i$ with the 
Coulomb functions in the external regions. This way all many-body matrix 
elements can be reduced to analytic forms plus one-dimensional integrals
\cite{Kamimura}.

The Siegert variational method uses a trial
function with purely outgoing asymptotics:
\begin{equation}
\Psi^t=\sum_{i=1}^Nc_i\varphi_i+c_{N+1}\phi^+.
\end{equation}
After the variation one arrives at  
a set of linear equations which is underdetermined (with $N+2$
unknowns:$\;c_1,c_2,$ $\dots,c_{N+1},E$). It has solutions, the bound 
states, only at discrete energies, where the determinant of the system 
of equations is zero. A matching technique similar to the scattering 
case can be used to calculate all matrix elements analytically. Using 
the resulting wave functions the reaction cross sections can be 
calculated. In order to get rid of the trivial exponential energy 
dependence of the cross sections, which comes from the Coulomb barrier 
penetration, we use the astrophysical $S$-factor 
\begin{equation}
S(E)=\sigma (E)E\exp{\Big [2\pi\eta (E)\Big ]}, \hskip 1cm
\eta (E)={{\mu Z_1Z_2e^2}\over{k\hbar^2}}.
\end{equation}

The model and methods described above have been used to study the $^7{\rm
Be}(p,\gamma){^8{\rm B}}$ and $^3{\rm He}({^3{\rm He}},2p){^4{\rm
He}}$ solar reactions in $^4{\rm He}+{^3{\rm He}}+p$ and $\{{^3{\rm
He}}+{^3{\rm He}},{^4{\rm He}}+p+p\}$ cluster models, respectively.

\section{{\bf $^{\bf 7}$Be({\bf p}$,\mbox{\boldmath $\gamma$}$)$^{\bf 
8}$B} cross section constrained by A=7 and 8 observables}
The most uncertain nuclear input parameter in standard solar
models is the low-energy $^7{\rm Be}(p,\gamma){^8{\rm B}}$ radiative
capture cross section. This reaction produces $^8$B in the sun, whose 
$\beta^+$ decay is the main source of the high-energy solar neutrinos. 
Many present and future solar neutrino detectors are sensitive mainly 
or exclusively to the $^8$B neutrinos. The predicted $^8$B neutrino flux 
is linearly proportional to $S_{17}$, the $^7{\rm Be}(p,\gamma){^8{\rm 
B}}$ astrophysical $S$ factor at solar energies ($E_{cm}=20$ keV). Thus, 
the value of $S_{17}$(20 keV) is a crucial input parameter in solar 
models. The six radiative capture measurements performed to date give, 
after extrapolations from higher energies, $S_{17}(0)$ between 15 eVb and 
40 eVb, with a weighted average of $22.4\pm 2.1$ eVb \cite{Johnson}, 
while a recent Coulomb dissociation measurement gives $S_{17}(0)=16.7\pm 
3.2$ eVb \cite{Motobayashi}. The theoretical predictions for $S_{17}(0)$ 
also have a huge uncertainty, as the various models give values between 
16 eVb and 30 eVb. 

The low solar energies mean that the reaction takes place well below the 
Coulomb barrier. In such cases the radiative capture cross section gets 
contributions almost exclusively from the external nuclear regions 
($r>6-8$ fm). At such distances the scattering- ($^7{\rm Be}+p$) and 
bound state ($^8$B) wave functions are fully determined, provided the 
scattering phase shifts and bound state asymptotic normalizations are 
known \cite{Xu}. At solar energies the phase shifts coincide with the 
(almost zero) 
hard sphere phase shifts, while the bound state wave function in the 
external region behaves like $\bar cW^+(kr)/r$, where $W^+$ is the 
Whittaker function and $\bar c$ is the asymptotic normalization. So the 
only unknown parameter that governs the $^7{\rm Be}(p,\gamma){^8{\rm 
B}}$ reaction at low energies is the $\bar c$ value. The $\bar c$ 
normalization depends mainly on the effective $^7$Be--$p$ interaction 
radius. A larger radius results in a lower Coulomb barrier, which leads 
to a higher tunneling probability into the external region, and hence to 
a higher cross section. We believe that the best way to constrain the 
potential radius is to study some key properties of the $A=7$ and 8 
nuclei \cite{ENS}. The observables that are most sensitive to the 
interaction radius are ``size'' type properties, for example, quadrupole 
moment, radius, Coulomb displacement energy \cite{Brown}, etc. 

\begin{figure}[tb]
\begin{center}
\leavevmode
\psfig{figure=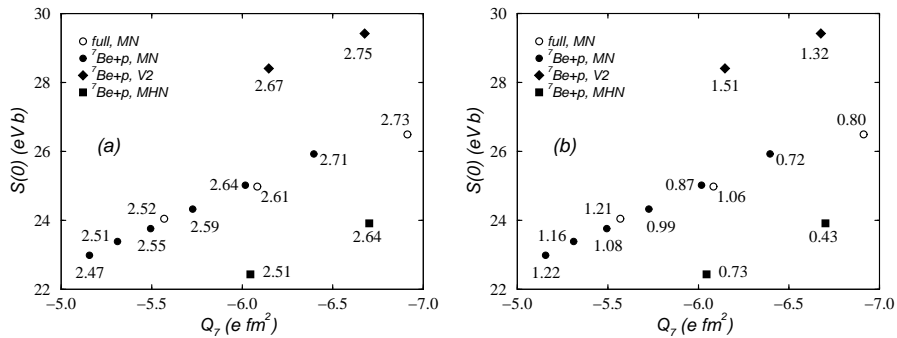,height=4.5cm}
\end{center}
\caption[]{Correlation between the zero-energy astrophysical $S$ factor
of the $^7{\rm Be}(p,\gamma){^8{\rm B}}$ reaction and the quadrupole 
moment of $^7$Be in our microscopic eight-body model. The results of 
several calculations, using various $N-N$ interactions and model spaces, 
are shown. Within one model space and interaction, the different results
come from different cluster sizes. The numbers in Fig.\ (a) are the
calculated $^8$B point-nucleon radii (in fm), while in Fig.\ (b) they are
the $r^2({^8{\rm B}})-r^2({^7{\rm Be}})$ values (in fm$^2$) for the 
various models. The phenomenological values are $r({^8{\rm B}})= 
2.50\pm0.04$ fm and $r^2({^8{\rm B}})-r^2({^7{\rm Be}})\approx0.9$ 
fm$^2$.}
\end{figure}

We use an eight-body three-cluster model which is variationally 
converged and virtually complete in the $^4{\rm He}+{^3{\rm He}}+p$ 
cluster model space \cite{PRC}. We find that the low-energy 
astrophysical $S$ factor is linearly correlated with the quadrupole 
moment of $^7$Be (Fig.\ 1).  This quantity, $Q_7$, has not been measured 
yet, but the model itself predicts it to be between $-6$ e$\;$fm$^2$ and 
$-7$ e$\;$fm$^2$. In addition to the $Q_7$ dependence, there is a 
sensitivity of $S_{17}(0)$ on the employed $N-N$ interaction, as shown 
in Fig.\ 1. We found that the MN interaction is the most self-consistent
in describing the $A=7$ and 8 nuclei and the $N+N$ systems. Thus, our 
model predicts $S_{17}(0)=25-26.5$ eVb. We mention, however, that the 
construction and use of other high-quality effective $N-N$ interactions 
would be desirable in order to check our findings. 

We have also tested how our model reproduces other ``size'' observables
\cite{TBP1}.
The quantity that is most sensitive to the effective $^7{\rm Be}-p$
interaction radius is $r^2({^8{\rm B}})-r^2({^7{\rm Be}})$. However, a
precise experimental determination of this quantity is very difficult.
Originally the $^7$Be and $^8$B radii were determined from interaction
cross sections by using Glauber-type models with uniform density
distributions for the nuclei \cite{Tanihata}. Recently a more precise
$^8$B radius was extracted from interaction cross section data by taking
into account the $^7{\rm Be}+p$ nature of $^8$B \cite{AlKhalili}. The
resulting point-nucleon radius is $r({^8{\rm B}})=2.50\pm0.04$ fm, 
and hence $r^2({^8{\rm B}})-r^2({^7{\rm Be}})\approx0.9$ fm$^2$. The
model still uses the Glauber result for the $^7$Be radius. A few-body
$^4{\rm He}+{^3{\rm He}}$ description of $^7$Be would probably slightly 
increase $r({^7{\rm Be}})$ and consequently $r({^8{\rm B}})$.

In Fig.\ 1 we give the $^8$B radii and $r^2({^8{\rm B}})-r^2({^7{\rm 
Be}})$, calculated for the various model spaces, interactions, etc. It
appears from Fig.\ 1(a) that the phenomenological $r({^8{\rm B}})$ value 
would suggest a $\approx$10$\%$ reduction in $S_{17}(0)$ relative to our 
MN prediction. However, it is interesting to note that increasing the 
model space (open circles) reduces the calculated $r({^8{\rm B}})$ at a 
given $Q_7$. If further model-space extensions resulted in the same 
behavior, then once again our MN results would be the most 
self-consistent. Fig.\ 1(b) shows that $r^2({^8{\rm B}})-r^2({^7{\rm 
Be}})$ seems to be already too small for both the MN and MHN 
interactions. Model-space extensions bring the MN results toward the 
phenomenological value. 
The $^8{\rm Li}-{^8{\rm B}}$ Coulomb displacement energy shows
similar behavior to the observables in Fig.\ 1 \cite{TBP1}.

\begin{figure}[tb]
\begin{center}
\leavevmode
\psfig{figure=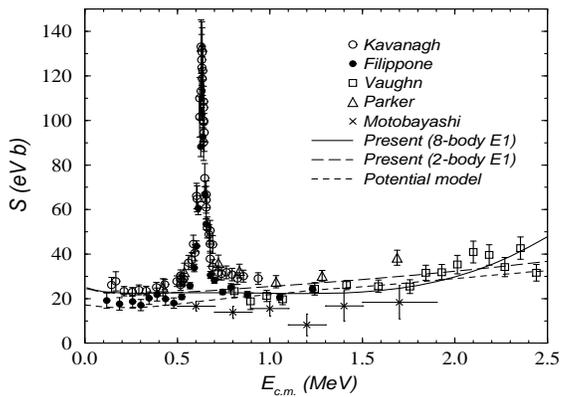,height=5.3cm}
\end{center}
\caption[]{Astrophysical $S$ factor for the $^7{\rm Be}(p,\gamma){^8{\rm
B}}$ reaction in our eight-body model. The symbols show the various 
experimental data (see \cite{PL} for references). The solid and 
long-dashed lines are the $E1$ components of the $S$ factors in our 
model with and without antisymmetrization in the electromagnetic 
transition matrix, respectively. The short-dashed line is the result of 
a typical potential model \cite{Xu}.}
\end{figure}

While the zero-energy cross section is sensitive only to the asymptotic
parts of the wave functions, with increasing energy the internal wave
functions become more and more important. The internal wave functions are
sensitive to effects like the exchange of the incoming proton with a
proton in $^7$Be, the excitation of the $^7$Be core by the incoming
proton, $^7$Be deformations, etc. The influence of these off-shell
effects on $S(E)$ was studied in our eight-body model \cite{PL}. We took
the $^7{\rm Be}-p$ relative motion wave functions coming from the cluster
model, and used them as if they came from a potential model. This way the
antisymmetrization, which is the biggest off-shell effect, was neglected
in the electromagnetic transition matrix.
One can see in Fig.\ 2 that the resulting $S$ factor (long dashed line)
has quite different energy dependence than the full microscopic $S$
factor (solid line). We also show an $S$ factor coming from a potential
model (short dashed line). This calculation demonstrates that there are
strong off-shell effects present in $^7{\rm Be}(p,\gamma){^8{\rm B}}$,
which have to be taken into account for any reliable extrapolation of
high-energy measurements.

All existing microscopic calculations for $^7{\rm Be}(p,\gamma){^8{\rm 
B}}$ use effective $N-N$ interactions with Gaussian shape. One can argue 
that using potentials with the correct Yukawa asymptotics would change 
$S(0)$. We studied this problem by using the MN interaction with Gaussian 
shape matched with a Yukawa tail \cite{TBP1}. We found that in the 
perturbative regime, where such a study makes sense, $S(0)$ is 
insensitive to the Yukawa tail, although the $A=7$ and 8 binding 
energies change significantly, and the overall strengths of the 
potentials have to be refitted.

Another interesting question is the effect of further extensions in the
model space by including, for example, $^6{\rm Li}+p+p$ and other
configurations in $^8$B. As an exploratory investigation, we studied the
effects of $^6{\rm Li}+N$ on $^7$Li and $^7$Be \cite{TBP1}. An
interesting result is that while $^7$Li is not affected by the $^6{\rm
Li}+n$ channel, some $^7$Be properties, like the quadrupole moment, is
strongly influenced by $^6{\rm Li}+p$. We observe a slight change also in
the energy dependence of the $^4{\rm He}({^3{\rm He}},\gamma){^7{\rm
Be}}$ $S$ factor if $^6{\rm Li}+p$ is included. Further studies are in 
progress.

\section{The {\bf $^{\bf 3}$He($^{\bf 3}$He,2p)$^{\bf 4}$He} reaction}
The $^3{\rm He}({^3{\rm He}},2p){^4{\rm He}}$ reaction competes with the 
$^7$Be producing branch of the solar p-p chain \cite{Bahcall}. Thus, it 
indirectly affects the $^7$Be and $^8$B neutrino fluxes. There are two 
interesting problems related to this reaction: i) a possible low-energy 
resonance in the cross section would suppress the high-energy solar 
neutrino fluxes \cite{Fowler}; ii) this is the only solar reaction whose 
cross section has been measured down to solar energies, and the effect of 
electron screening is still not fully understood \cite{LUNA}. We studied 
these problems in a six-body $\{{^3{\rm He}}+{^3{\rm He}},{^4{\rm 
He}}+p+p\}$ cluster model \cite{TBP2}. 

If there is a resonance in the $^3{\rm He}({^3{\rm He}},2p){^4{\rm He}}$ 
reaction cross section, then it comes from either ${^3{\rm He}}+{^3{\rm 
He}}$ or $^4{\rm He}+p+p$. Interestingly, the second case is easier to
study despite its three-body nature. We searched for high-lying narrow
resonances in $^4{\rm He}+p+p$ using the complex scaling method that can
handle the three-body Coulomb asymptotics correctly \cite{3br,TBP2}. We
found no such states. The ${^3{\rm He}}+{^3{\rm He}}$ channel is more
difficult and more interesting. We mention here only one interesting
feature. The ${^3{\rm He}}+{^3{\rm He}}$ system is similar in many
respects to the $n+n$ system, which has a virtual $0^+$ state with
negative energy and negative imaginary wave number. If there is a virtual
state present in ${^3{\rm He}}+{^3{\rm He}}$ then it would result in a 
$^3{\rm He}({^3{\rm He}},2p){^4{\rm He}}$ cross section that is singular 
at the (unphysical) negative ${^3{\rm He}}+{^3{\rm He}}$ pole energy, and 
has a $1/E$ energy dependence at low positive energies. We show the 
effect of such a hypothetical state in Fig.\ 3. The observed rise in the 
measured cross section (relative to the bare cross section) is attributed 
to the effect of electron screening. One can see that a virtual state 
could mimic a similar behavior. We have searched for virtual states in 
${^3{\rm He}}+{^3{\rm He}}$ in a simple cluster model, and found none so
far. Further studies in a more realistic model are in progress
\cite{TBP2}.

\begin{figure}[tb]
\begin{center}
\leavevmode
\psfig{figure=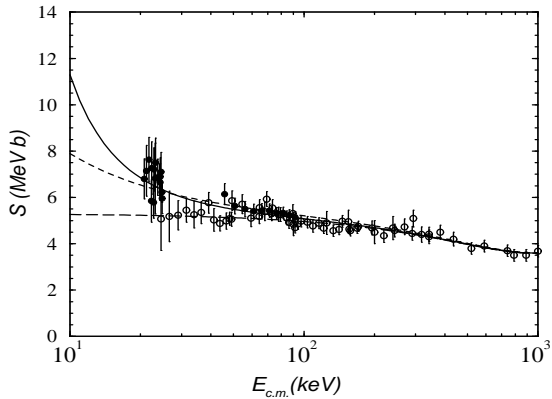,height=5.3cm}
\end{center}
\caption[]{Astrophysical $S$ factor for the $^3{\rm He}({^3{\rm 
He}},2p){^4{\rm He}}$ reaction. The data points are from \cite{LUNA} and
references therein. The solid line is the fitted raw cross section, 
while the long dashed line is the bare cross section, determined by 
removing the electron screening effect with $U=323$ eV screening 
potential. The short dashed line shows the effect of a hypothetical 
zero-energy virtual state on the bare cross section. The strength of 
the virtual state is artificially set to 5 keV.}
\end{figure}

The understanding of the energy dependence of the cross section is also
important for the study of electron screening effects. The screening
potential, extracted in \cite{LUNA}, seems to be larger than predicted by
theory. We studied the $^3{\rm He}({^3{\rm He}},2p){^4{\rm He}}$ and the 
mirror $^3{\rm H}({^3{\rm H}},2n){^4{\rm He}}$ reactions in large-space 
cluster models in the continuum discretized coupled channel approximation
\cite{TBP2}. The results are in Fig.\ 5. We observe a good general
agreement with the data in both the absolute normalization and shape.
However, a marked disagreement exists at very low energies in 
$^3{\rm H}({^3{\rm H}},2n){^4{\rm He}}$ between our result and the very
precise Los Alamos data. Further investigation to understand this
discrepancy is in progress.

\begin{figure}[tb]
\begin{center}
\leavevmode
\psfig{figure=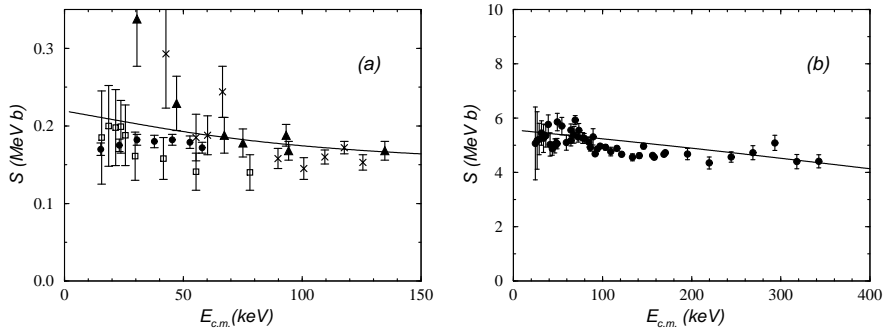,height=4.43cm}
\end{center}
\caption[]{Astrophysical $S$ factor for the (a) $^3{\rm H}({^3{\rm 
H}},2n){^4{\rm He}}$ and (b) $^3{\rm He}({^3{\rm He}},2p){^4{\rm He}}$ 
reactions. In Fig.\ (b) only the data of Krauss are shown (cf.\ 
\cite{LUNA}). The curves come from our six-body calculations in a 
continuum discretized coupled channel approximation.}
\end{figure}
\section{Conclusion}We have studied the $^7{\rm Be}(p,\gamma){^8{\rm 
B}}$, and $^3{\rm He}({^3{\rm He}},2p){^4{\rm He}}$ solar nuclear
reactions in microscopic cluster models. 

Our $S_{17}(0)=25-26.5$ eVb astrophysical $S$ factor is slightly higher
than, but consistent with the value \cite{Johnson} $22.4\pm2.1$ eVb 
currently used in standard solar models. Our results show that certain
$^7$Be and $^8$B observables, like radius and quadrupole moment,
can establish a region of the possible values of $S_{17}(0)$. Our model
shows that a small value of $S_{17}(0)$ is rather unlikely, unless
some very important ingredient is missing in our approach. Currently we
have no candidate for such a missing element. We would like to emphasize
the need for further experiments by using both the radiative capture
technique \cite{Hammache} and Coulomb dissociation \cite{CD}. A
measurement of the $^7$Be quadrupole moment and a precise extraction of
the $^7$Be radius would also be very beneficial.

Our calculations for $^3{\rm He}({^3{\rm He}},2p){^4{\rm He}}$ show no
resonances or virtual states either in the $^3{\rm He}+{^3{\rm He}}$ or 
in the $^4{\rm He}+p+p$ channels. Further studies of $^3{\rm He}+{^3{\rm 
He}}$ in more realistic models are necessary. We 
calculated the $^3{\rm He}({^3{\rm He}},2p){^4{\rm He}}$ and $^3{\rm 
H}({^3{\rm H}},2n){^4{\rm He}}$ cross sections and found a good general 
agreement with existing data. 

Solar model independent analyses show that the measured solar neutrino
rates cannot be reproduced by arbitrarily changing the normalizations of
the neutrino spectra \cite{Heeger} (the shapes of the spectra are
currently beyond any doubt). Thus, the solar fusion rates themselves, 
which can change only the absolute normalizations of the fluxes, cannot
solve the solar neutrino problem. In fact we have more efficient ways
within the standard model to change the absolute normalizations, than
nuclear physics. For instance, by slightly deviating from the
Maxwell-Boltzmann thermal statistics one can cause as large changes in
the various neutrino fluxes as those that constitute the solar neutrino 
problem itself \cite{Maxwell}. However, in order to have a solution of 
the solar neutrino problem within the standard model, we would still need 
a yet unknown mechanism that could cause spectrum distortion, like the 
MSW neutrino oscillation in a non-standard model \cite{Bahcall}.

\section*{Acknowledgments}
This work was performed under the auspices of the U.S.\ Department of
Energy. I am grateful to Prof.\ Karlheinz Langanke for many useful
discussions. This work was partly supported by the Danish Research
Council and the Theoretical Astrophysics Center during my stay at Aarhus
University.

\end{document}